RESEARCH ARTICLE

# Using a neural network approach and starspots dependent models to predict effective temperatures and ages of young stars

Marco Tarantino 1*, Loredana Prisinzano2, Nicoletta D'Angelo1, Francesco Damiani2, Giada Adelfio1

1 University of Palermo, Department of Economics, Business and Statistics, Palermo, Italy,
2 INAF-Osservatorio Astronomico di Palermo, Palermo, Italy

* marco.tarantino01@unipa.it

## Abstract

This study presents a statistical approach to accurately predict the effective temperatures of pre-main sequence stars, which are necessary for determining stellar ages using the isochrone methodology and cutting-age starspots-dependent models. By training a Neural Network model on high-quality spectroscopic temperatures from the Gaia-ESO Survey as the response variable, and using photometric data from Gaia DR3 and 2MASS catalogs as explanatory variables, we implemented a methodology to accurately derive the effective temperatures of much larger populations of stars for which only photometric data are available. The model demonstrated robust performance for low-mass stars with temperatures below 7 000 K, including young stars, the primary focus of this work. Predicted temperatures were employed to construct Hertzsprung-Russell diagrams and to predict stellar ages of different young clusters and star forming regions through isochrone interpolation, achieving excellent agreement with spectroscopic-based ages and literature values derived from model-independent methods like lithium equivalent widths. The inclusion of starspot evolutionary models improved the age predictions, providing a more accurate description of stellar properties. Additionally, the results regarding the effective temperature and age predictions of the young clusters provide evidence for intrinsic age spreads in the youngest clusters, suggesting multiple formation events over time.

## Introduction

Understanding the physical mechanisms involved in the star formation (SF) is still an open question, since it concerns not only the complex physics of individual newborn stars at ages $\lesssim 10$ Myr (activity, accretion or circumstellar disks), but also the context in which they are found. In fact, it's by now widely recognised that stars form in clusters within giant molecular clouds [1]. SF arises from gravitational instability within these high density and turbulent molecular clouds, leading to the collapse of archives. The tables can be retrieved from: https://doi.eso.org/10.18727/archive/25. We built a repository with the codes and data used in the analysis which can be visited with the following DOI: https://doi.org/10.5281/zenodo.17159340.

**Funding:** The research work of Nicoletta D'Angelo and Giada Adelfio was supported by the Targeted Research Funds 2025 (FFR 2025) of the University of Palermo (Italy) and the European Union - NextGenerationEU, Mission 4, Component 2, in the framework of the GRINS - Growing Resilient, INclusive and Sustainable project (GRINS PE00000018 – CUP C93C22005270001). MUR- PRIN 2022: Spatio-temporal Functional Marked Point Processes for probabilistic forecasting of earthquakes 2022BN7CJP. P. I. Giada Adelfio. CUP B53C24006340006.

**Competing interests:** The authors have declared that no competing interests exist.

the cloud and subsequent SF [2,3]. To describe such process, two different SF models have been proposed. One is the bottom-up or hierarchical model in which high levels of SF efficiency are observed in clumps or filaments, leading to the formation of subclusters that merge (e.g. [4,5]). Other authors (e.g. [6]) have suggested a top-down or monolithic model, in which the cluster formation occurs in a rapid and efficient single SF episode.

Different mechanisms can be responsible for increasing the density of the clouds: stellar feedback from a supernova or massive star winds [7], dynamic feedback from cloud–cloud collisions [8], or ionisation of hydrogen gas by massive stars. Such mechanisms are thought to trigger SF but, also, increase the duration of the process by forming different generations of stars within a single cluster (e.g. [9,10]). This is an open issue in a large context, including evidences of multiple stellar generations in Globular Clusters, where SF persisted for several million years with a lack of feedback in young starburst [11]. Many observational studies aim at finding evidence supporting one theoretical scenario over another, or evidence of positive feedback, such as triggered SF. These analyses are often based on observations of molecular gas (e.g. [12]).

An alternative approach is based on the statistical analysis of the youngest stellar populations in star forming regions (SFR). It consists in empirically constraining the timescales involved with cluster formation. Such approach suffers from the well known problem of determining accurate stellar ages (e.g. [13–15]). One of the most important challenges in the field of stellar astrophysics is defining an absolute age scale for pre-main sequence stars (PMS), as it is crucial not only for understanding cluster formation processes, but also to constrain the time-scale involved in their evolution, in the mechanisms of circumstellar material dispersal, and planet formation [16].

PMS stellar ages are generally derived with two different methods. One is based on the photospheric lithium (e.g. [17,18]) which is the least model-dependent one. However, it requires many spectroscopic observations which are telescope time-demanding. Even in the epoch of multi-object spectrographs (e.g. 4MOST [19], WEAVE [20]), such observations can be acquired for a limited fraction of relatively bright stars. The other method is the classical Hertzsprung–Russell (HR) diagram method, which relies on comparing theoretical isochrones with observational stellar features. However, the method is sensitive to various factors, including the treatment of convection, magnetic fields, star spots, rotation [21,22], as well as the accurate derivation of extinction and distances, the reddening laws and the conversions between effective temperatures and colors [23].

The *Gaia* mission of the European Space Agency (ESA) [24] is revolutionizing our ability to address many of these challenges. The details and survey characteristics of the latest data release (DR3) are outlined in Vallenari et al. (2023) [25]. *Gaia* is providing astrometry, i.e. positions, parallaxes and proper motions with unprecedented precision, down to G$\lesssim$20 mag [26], as well as an unprecedented accurate and all-sky homogeneous photometry in the bands G, $G_{BP}$ and $G_{RP}$ [27]. Such data, obtained for over 1.5 billion stars, are enabling for the first time significant advances in our ability to derive precise membership (e.g. [28–30]) and accurate stellar distances [31].

However, accurately positioning stars on the HR diagram requires the determination of the effective temperatures, which is crucial to derive also individual stellar extinction values, especially for YSOs, which are often affected by spatially variable extinction. Existing reddening maps [32] are insufficient for this purpose, as they do not account for local variations in extinction.

*Gaia* provided also spectroscopic stellar information from the Radial Velocity Spectrometer (RVS) data [33] but such parametrisation is limited to about 5.6 million bright stars down to G~16. Stellar information have also been derived using *Gaia* BP/RP spectra, which extend down to G~19. However, these values are significantly affected by a temperature-extinction degeneracy, particularly for dwarfs, and should be used with caution for the low-mass stars [34].

The *Gaia*-ESO Survey (GES) is an European Southern Observatory (ESO) Public spectroscopic survey (ESO Programme IDs: 188.B-3002, 193.B-0936, 197.B-1074; PIs: Gerry Gilmore & Sofia Randich) consisting of 340 nights of observations, carried out between 2012 and 2018 [35–38]. It covers all the major components of the Milky Way, namely Bulge, Halo, Thick Disc and Thin Disc populations, open clusters and SFRs [39]. More than 110 000 stars with G≲19 were observed with the Fibre Large Array Multi Element Spectrograph (FLAMES) of the 8.2 m-diameter Very Large Telescope (VLT) of ESO. These spectroscopic observations provide important information such as the measurements of the effective temperature, gravity, chemical abundances and radial velocity of the observed stars.

In the last few years, also other astronomical datasets are facing a rapid growth in size and complexity, thus introducing Astronomy to the era of big data science. This growth is a result of past, ongoing, and future photometric and spectroscopic surveys, such as 2MASS [40], the Sloan Digital Sky Survey [41], Pan-STARRS [42], Rubin LSST [43], VVV [44], AllWISE [45], LAMOST [46] and APOGEE [47]. Considering this accelerated growth in available astrophysical data, it is increasing the necessity to use statistical methods for extracting, exploring and modeling these data in order to detect, characterise and classify astrophysical objects [48]. Machine learning (ML) algorithms have gained increasing popularity among astronomers, and are widely used for a variety of tasks (e.g. [49–52]).

The aim of this work is to apply statistical techniques in order to predict as accurately as possible the effective temperature and the ages of pre-main sequence (PMS) stars, using the data extracted from the Gaia DR3, GES, and 2MASS catalogs. In particular, the GES catalog provides spectroscopic temperature measurements, which are highly accurate but only available for targeted and relatively bright objects, meaning they are limited to smaller samples of stars. On the other hand, the Gaia DR3 and 2MASS surveys provide photometric information for a much larger set of stars, including colors and magnitudes in various filters, which can typically be used to obtain photometric temperature predictions. However, these photometric values can be significantly biased by the extinction effects, which are especially problematic for young stellar objects (YSOs), where extinction can vary spatially. In this work, we propose a statistical method able to understand the relationship between the photometric variables and the spectroscopic temperature values, in order to predict the effective temperature for all the pre main sequence stars for which the photometric information are available.

For this purpose, we propose a Neural Network model. The model has the potentiality to be applied to generic photometric datasets, based on Gaia DR3 and 2MASS surveys, including hundreds of millions of stars for which spectroscopic measurements are not available. We can predict effective temperatures for all these stars and subsequently derive extinction values, which are crucial for determining accurate stellar luminosities. These quantities are essential for placing stars in the HR diagram and compare their position with state-of-the-art theoretical isochrones to accurately predict stellar ages. To validate our model, we selected young clusters for which accurate membership has been derived [53]. We compare our results with very recent age determinations obtained for the relevant clusters to infer our conclusion on the cluster age spread.

The analysis presented in this work was conducted using both Python [54] and R [55]. All the codes and analysis are available from the authors.

The article is structured as follows. In *Data and Methods* we describe all the data extraction and selection procedures starting from the Gaia, GES and 2MASS archives to obtain the final dataset for the analysis of interest. We introduce the statistical methods used to perform feature selection and to predict the YSOs effective temperatures. Additionally, we describe the isochrone method, the starspot evolutionary models and the statistical procedures used to obtain stellar age predictions and the associated uncertainties, necessary to validate our temperature predictions. All the results of temperature and age predictions are presented in *Results*. In the *Discussion*, we compare our results with previous studies and provide an interpretation of their implications. In the final section, we draw our conclusions and suggest potential avenues for future research.

## Data and methods

### Input dataset

The GES catalogue used in this work is based on the data provided by the consortium with last internal data release GES iDR6 [56]. The catalogue is the result of the work of a large consortium structured in 19 Working Group (WG). The general homogenization of the astrophysical information and elemental abundances has been carried out by WG15. A description of the general work flow is given in Hourihane et al. [57]. The initial GES table includes features of 114 917 stars. From this table, stars without effective temperature values were excluded, thus reducing the set of stars to 89 783. We used the *Gaia* DR3 `source_id`, provided in the GES iDR6 data products, to retrieve the *Gaia* and the 2MASS photometry from the ESA's Gaia archive [58], using the official cross-match between Gaia and 2MASS [59]. The data were retrieved using the Astronomical Data Query Language (ADQL).

As a final selection is critical for the ongoing analysis, only stars with a GES spectroscopic temperature below 8 000 K were included, aiming to discard stars with extreme values (the maximum temperature observed in the GES dataset is 45 000 K), that are behind the aim of this work. This selection results in a reduced dataset of 84 884 stars.

Distances of the targets included in our dataset have been taken from Bailer-Jones et al. (2021) [60]. In particular, we used the photogeometric distances, based on the *Gaia* parallax, with a direction-dependent prior on distance, and the color and apparent magnitude of a star. As for the distribution of distances, relative to the selected stars, it presents a positive skewness, with a value of the relative estimate equal to $s = 1.475$, while the kurtosis value turns out to be $k = 2.958$, almost equal to the value of a classic Gaussian ($k = 3$). The 90% of the stars shows a distance value between 449 pc and 7 013 pc, the average distance of these stars is 2 709 pc, with an observed standard deviation of 2 022 pc. There are some very distant stars, in fact more than four hundred stars exceed a distance of 10 kpc. Despite these high distances, we decided to retain these stars based on two primary considerations. Including a larger amount of data allows the model to learn more effectively, enabling it to capture the relationships between the explanatory variables, or covariates, and the response variable more comprehensively. Removing these stars would reduce the dataset's diversity, potentially weakening the model's capacity to generalize. Furthermore, a preliminary exploratory analysis of these distant stars showed that their covariates' values were not anomalous. Therefore, we concluded that retaining these stars, even at high distances, would not affect the robustness or accuracy of the model.

Since our aim is to develop a model designed to predict effective temperatures with accuracy similar to the spectroscopic one, we have integrated the input dataset by including two additional stellar information, the 'input' extinction and the 'input' effective temperatures, preliminarily predicted using only photometric data, as detailed below.

**Input extinction values from reddening maps.** Preliminary extinction values were obtained using the 3D reddening maps by Vergely et al. (2022) [61]. These extinction density maps are given for three different volumes and maximum spatial resolutions. To make use of the best resolution, we employed all three maps depending on the star's distance. Specifically, we used the map with the highest resolution (10 pc) for stars with distances less than 1.5 kpc, as it is defined within the volume of 3 000 pc × 3 000 pc × 800 pc around the Sun. For stars with distances between 1.5 kpc and 3 kpc, we utilized the map with intermediate resolution (25 pc), defined within the volume of 6 000 pc × 6 000 pc × 800 pc around the Sun. Finally, for stars with distances greater than 3 kpc, we employed the map with the lowest resolution (50 pc), which is defined within the maximum volume of 10 000 pc × 10 000 pc × 800 pc around the Sun.

We used the Python `RegularGridInterpolator` function to interpolate the reddening value for each star at a given distance $d$ and Galactic coordinates ($l,b$), converted to Cartesian coordinates (X, Y, Z). The algorithm takes two arguments as input: the tuple of arrays defining the spatial grid of the map (one for each dimension) and the multidimensional array of reddening values corresponding to the grid. Once the interpolator object is created, it can be used to obtain interpolated values at any point (X, Y, Z). To take into account the possible spatial dust variability, typical of star forming regions, we used the 'nearest' method, which is more suitable for representing discontinuities and variations in density.

The nearest-neighbor interpolation based on the maps can be unreliable beyond 3 kpc where the resolution map is very low (50 pc). Our sample of 84 884 includes 29 298 stars with distance $\geq$ 3 kpc. To assess if the preliminary extinction values derived for stars with distances >3 kpc can introduce potential biases in deriving our final parameters (effective temperatures and ages), we compared the $A_G$ values from the Vergely ($A_{G_{phot}}$) maps with the $A_G$ values derived from the GES spectroscopic effective temperatures ($A_{G_{GES}}$), by distinguishing the subsample with distance <3 Kpc and the one with distance $\geq$ 3 Kpc. Such comparisons show that the computed MAD between $A_{G_{phot}}$ and $A_{G_{GES}}$ for stars with distance $\geq$ 3 kpc is 0.273, while for stars with distance <3 kpc, such dispersion is 0.13. This means that the results could be worse for distant stars, as will be discussed in Section *Results*.

The $A_V$ values derived in this way were converted into extinctions in the Gaia bands $G$, $G_{BP}$, $G_{RP}$ and in the near-infrared 2MASS bands (J, H, K), by using the recommended *Gaia* EDR3 reddening laws [62], obtained following the method presented in Danielski et al. (2018). [63] In particular, we used the formula derived as a function of the star's intrinsic colour $G_{BP} - G_{RP}$, assuming, as a first approximation, that intrinsic colors correspond to the observed ones. Once the absorption values in the BP and RP bands were computed, we derived the intrinsic ($G_{BP} - G_{RP}$)$_0$ colors and then re-calculated the absorption values in the various bands. The second iteration resulted in reddening values $E(G_{BP} - G_{RP})$ that differed by less than 1 milli-magnitudes from the previous values. Therefore, a single iteration was sufficient to achieve convergence of the values.

**Input photometric temperatures.** Intrinsic, i.e. absorption-corrected colours are related to the effective temperature of the stars. In this sub-section we describe how, using the provisional absorption values obtained as described in the previous section, and an appropriate empirical relation, a provisional prediction of the effective temperature can be derived, which we define here as *photometric effective temperature* $T_{eff}^{phot}$. As described in Pancino et al. (2022), [50] it is considered as an additional explanatory variable to be included in the subsequently applied statistical models, which is useful for the final prediction of the effective temperature.

To derive the values of $T_{eff}^{phot}$, we used an updated version of the polynomial relation between the *Gaia* DR3 colours ($G_{BP} - G_{RP}$)$_0$ and the effective temperatures, derived as in Montalto et al (2021) [64].

## Feature selection

Given the amount of potential data available, the challenge lies in selecting the most informative explanatory variables, or features, to obtain accurate temperature predictions. Hence, a crucial step is feature selection [65], which reduces the dimensionality of the problem, necessary to avoid multicollinearity, by focusing on the most relevant explanatories for the analysis.

Multicollinearity refers to the linear relationship among two or more variables, which also means lack of orthogonality among them. In more technical terms, multicollinearity occurs if $p$ vectors lie in a subspace of dimension less than $p$. This is the definition of exact multicollinearity or exact linear dependence. It is not necessary for multicollinearity to be exact in order to cause a problem, it is enough to have $p$ variables strongly correlated [66].

High multicollinearity significantly affects the performance of parametric models. The high correlation among the covariates $x_j$ leads to a loss of model power, as the standard error of model parameters are overestimated and consequently the significance level of these parameters is biased [67].

In contrast, non-parametric models are less sensitive to multicollinearity due to their architecture, which can redistribute importance across correlated predictors. Neural networks, for instance, are non-parametric models that use layers and non-linear activation functions, which allow them to learn representations of the data in a way that mitigates the direct impact of correlated inputs [68]. However, multicollinearity can still cause problems also in non-parametric models, such as increasing the risk of overfitting, where the model captures noise rather than the underlying signal, and increasing computational complexity, as redundant information leads to inefficiencies in model training and convergence.

Various techniques have been developed to handle multicollinearity. Some of the most used in statistics are regularization methods, such as Lasso [69] and Ridge [70,71] regressions, which introduce penalties in the model, shrinking the coefficients related to less important covariates and reducing the impact of correlated predictors. Ridge regression aims to minimise the sum of squares of the errors by adding a proportional term to the sum of squares of the coefficients multiplied by a regularization parameter $\lambda$ (L2 penalty). The higher the penalization, the more the estimated coefficients are shrunk toward zero, without ever reaching exactly zero value, which would eliminate them from the model. On the other hand, Lasso regression uses a term equal to the sum of the absolute values of the coefficients multiplied by the regularization parameter $\lambda$ as penalty (L1 penalty). Lasso favours sparsification of the model, since the estimated coefficients can be shrunk to exactly zero, cancelling out the associated covariates and thus leading to direct feature selection.

In our work, we decided to perform feature selection through regularization techniques. In particular, we adopted the Elastic Net regression [72]. This regularization method can be seen as a combination of the Ridge and Lasso regressions, since simultaneously does automatic variable selection, as the Lasso regression, and continuous shrinkage, as the Ridge regression. Let $y_i$ be the response variable value related to the $i$-th observation, in our case it represents the effective temperature of the $i$-th star, $x_{ij}$ the value of the $j$-th covariate, i.e. a color, magnitude or other every possible explanatory variable observed for the $i$-th star, and $\theta$ the vector of unknown $p + 1$ parameters that we want to estimate, where $\theta_0$ represents the intercept of the model, i.e. the predicted value of the response variable when all the covariates are equal to zero, while each $\theta_j$ represents the effect on the response variable given by a unitary variation of the value of the respective $j$-th covariate. The Elastic Net regression is applied to obtain an estimation of $\theta$ by minimising the following function:

$$\min_{\theta} \left\{ \sum_{i=1}^{n} \left( y_i - \theta_0 - \sum_{j=1}^{p} \theta_j x_{ij} \right)^2 + \alpha \sum_{j=1}^{p} |\theta_j| + (1-\alpha) \sum_{j=1}^{p} \theta_j^2 \right\}, \tag{1}$$

where $n$ is the total number of observations and $p$ the number of covariates, while $\alpha$ corresponds to the regularization parameter, which takes values in [0,1]. The function $\alpha \sum_{j=1}^{p} |\theta_j| + (1-\alpha) \sum_{j=1}^{p} \theta_j^2$ is called Elastic Net penalty, which is a convex combination of the Lasso and Ridge penalty. If $\alpha = 0$, the Elastic Net becomes simple Ridge regression, while for $\alpha = 1$ it becomes a simple Lasso regression. The best value of $\alpha$ is found by the LARS-EN algorithm, which was implemented in Zou et al. (2005) [72], based on the LARS algorithm proposed in Efron et al. (2004) [73].

**Final dataset**

Referring to the information of the Gaia and 2MASS surveys, we have extracted six apparent magnitudes in G, BP, RP Gaia filters and J, H, K 2MASS filters and their respective errors, the galactic coordinates $l$ and $b$, the photo-geometric distances and their errors for all the 84 884 stars. Starting from these covariates, we have calculated different colors, such as $G-G_{BP}$, $G-G_{RP}$ and $G_{BP}-G_{RP}$ for Gaia and $J-H$, $H-K$ and $J-K$ for 2MASS. Also, we computed the absolute magnitudes as:

$$M_F = m_F - 5 \, log(d) + 5 - A_F \tag{2}$$

where $m_F$ and $M_F$ represent, respectively, the apparent and the absolute magnitudes for a given photometric filter $F$, $d$ stands for the star distance, and $A_F$ is the extinction in that filter.

These covariates are strongly correlated, since both magnitudes and colors are quantities related to the luminosity of the stars. In particular, colors are linear combinations of magnitudes and absolute magnitude values strongly depend on apparent magnitudes. Consequently, including all the colors, apparent magnitudes and absolute magnitude in the model, can cause a multicollinearity problem. The Elastic Net helps us to select a restricted number of covariates, reducing the model complexity and overcoming the multicollinearity. The set of selected features includes:

- the absolute magnitude in the Gaia *BP* filter, $(M_{BP})_0$;
- the apparent magnitude in the 2MASS *k* filter, $(m_k)_0$;
- the *Gaia* $(G - G_{RP})_0$, $(G_{BP} - G_{RP})_0$ colors;
- the $(J - H)_0$ and $(H - K)_0$ colours from 2MASS;
- the input photometric temperature, $T_{eff}^{phot}$.

Magnitudes and colors are preliminarily corrected for extinction, as described before. Fig 1 shows the correlations among the explanatory variables and their correlations with the response, i.e. the spectroscopic effective temperature ($T_{eff}^{GES}$). Some of the selected features still have strong internal correlations, for instance, the colors $G$-$G_{RP}$ and $G_{BP}$-$G_{RP}$, and the $T_{eff}^{phot}$. However, these explanatory variables are the ones which show the highest correlations with the response.

To verify that the selected set of features is the best one among all the possible sets, we conducted additional tests using alternative combinations of magnitudes and colors to predict the effective temperature by performing the Neural Network model, which is described in the following sub-section. Specifically, given that the $(M_{BP})_0$ magnitude is known to suffer from larger photometric errors compared to other magnitudes such as $(M_G)_0$ or $(M_{RP})_0$, we tested feature sets that excluded $(M_{BP})_0$ in favour of other available magnitudes. Despite these tests, we found that none of these alternative combinations provided a level of agreement with the spectroscopic temperature values comparable to the one achieved with the inclusion of $(M_{BP})_0$. This result highlights the contribution of $(M_{BP})_0$ to the model, even if its higher associated uncertainties. For this reason, we decided to keep $(M_{BP})_0$ as a covariate in the final model configuration, confirming the features set selected by the Elastic Net procedure.

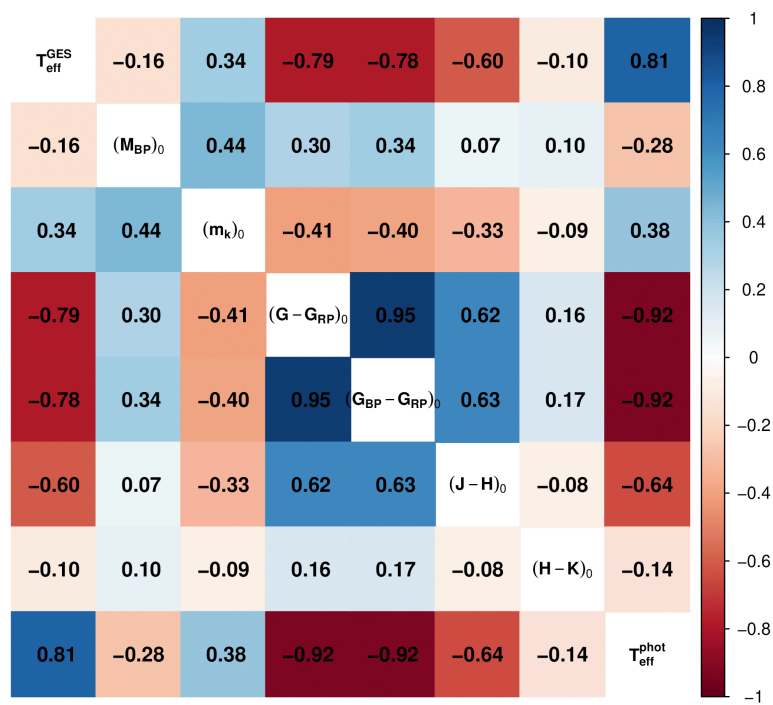

**Fig 1. Corrplot of the selected features.** The colour scale indicates the level of correlation between the two related variables. Magnitudes and colors are corrected for extinction.

**Neural network for temperature prediction**

In this section we introduce the statistical methods applied to the data to obtain predictions of the effective temperature. The complexity of the analysis in this study is given by the data, which consist of more than eighty thousand stars, each characterized by eight covariates that were selected through the Elastic Net regression. Despite this selection process, certain variables remain strongly correlated, as illustrated in Fig 1.

Neural Networks (NN) emerge as a promising approach to model these data, as it is a non-parametric approach able to handle complex and highly correlated data, due to its ability to learn non-linear relationships and interactions between features automatically. Unlike other non-parametric methods, NN can be adapted to intricate patterns in the data through its layered architecture, which progressively extracts higher-level representations.

The results of the NN approach are compared with benchmark models such as Random Forest [74,75] and Support Vector Machines [76], which are recognised as state-of-the-art statistical techniques for non-parametric regression problems.

To apply the Neural Network approach, several hyperparameters must be defined, because the structure, training process, and performance of the model depend on them. One of the most important choices is the number of layers in the network, which determines its depth. More layers can allow the model to capture increasingly complex relationships in the data, but they also increase the computational time and the possibility of overfitting if not properly managed. Along with the number of layers, the layer size must be specified, which refers to the number of neurons in each hidden layer as well as the output layer. Too few neurons may underfit, while too many can lead to overfitting and computational inefficiencies. Another important choice is the optimizer, which controls the algorithm used to minimise the error during training. Common choices include Adam, Stochastic Gradient Descent, and RMSprop, each of which has distinct properties in terms of convergence speed and suitability for different types of data [77]. It is also necessary to define the error function to be optimised. The batch size determines how many samples are processed in each training step. Smaller batch sizes can make the training process more stable and accurate, but they require more memory and can be slower. Lastly, the epochs define the total number of passes the model makes through the entire training dataset. More epochs allow the model to learn deeper, but it also increases the risk of overfitting if the model starts to memorise the training data instead of generalising from it. There are different methods to prevent overfitting. The most common are the Early Stopping [78], the Dropout [79] and the K-Fold Cross-Validation (KFold) technique [80]. We adopted this last method, which consists in dividing the training dataset into $k$ folds, where each fold is used as validation set exactly once, while the remaining folds form the training set. By using this approach, the model's performance is evaluated each time on different subsets of the data, reducing the risk of overfitting to a single validation set.

We compared the results of different architectures. The architecture of the Neural Network model that gives the best results consists in five hidden layers with progressively decreasing number of neurons: 128 neurons in the first hidden layer, followed by 64, 32, 16, and 8 neurons in the subsequent layers. Each of these layers uses the ReLU (Rectified Linear Unit) activation function $f(x) = max(0, x)$, a common choice due to its simplicity and efficiency in the training phase of Neural Networks. The output layer contains 1 neuron with a linear activation function, appropriate for regression goals where the aim is to predict a continuous target variable, such the effective temperature of the stars. The selected optimizer for model training is the Adam, while we set the mean absolute error as cost function to be optimised. We impose fifty epochs to train the model and batch size equal to eight. To prevent the model from overfitting, the K-Fold Cross-Validation (KFold) technique is used, with $k = 10$. The NN is trained on the training set, given by the 80% of the stars of our final dataset (67 904 stars). Considering the KFold methodology, this training set is divided into ten sets of $\approx$ 6 790 stars, where, at each iteration, nine of them are effectively used as training (61 114 stars) and one as validation set. The performance of the NN are then tested on the test set, made up by the remaining 20% of the stars of the final dataset (16 980 stars). A scheme of the architecture selected for our Neural Network is shown in Fig 2.

Once the method to predict the temperature is set, it is important to associate an error measure to each prediction. While an estimate may be accurate, without an error measure we don't know how reliable it is. The standard errors help to understand the precision of the estimates and provide a measure of their variability. To estimate the standard errors, we take into account the bootstrap methodology [81]. Bootstrap is a statistical technique used to estimate the variability of predictions obtained through statistical models and to calculate standard errors, confidence intervals and other measures of uncertainty without making assumptions about the distribution of the data. The method consists in repeatedly sampling observations, with replacement, from the available data, creating different versions of the training set and training the model on each of these versions, obtaining the corresponding predictions. By repeating this process $n_b$ times, we obtain $n_b$ temperature predictions for the same star. From the empirical distribution of these $n_b$ temperature predictions, we can therefore evaluate their variability. The higher is $n_b$, the more robust will the variability estimation of the predictions be. We will use the bootstrap procedure on the chosen NN, considering $n_b = 100$.

A good model, theoretically, should have a high descriptive ability on the data on which it is trained and, simultaneously, a high generalization ability on the test data. In other words, it is hoped that prediction errors will be low on both the training and the test sets. In order to assess the predictive ability of the model, the predicted values of the response ($\hat{y}_i$) are compared with its observed values ($y_i$). In this work, to assess the model's predictive ability we computed the Median Absolute Deviation (MAD) on the model's residuals. We use this quantity to compare the results of the NN with the Random Forest and the Support Vector Machine approaches, which are trained on the same training set, using also the KFold methodology.

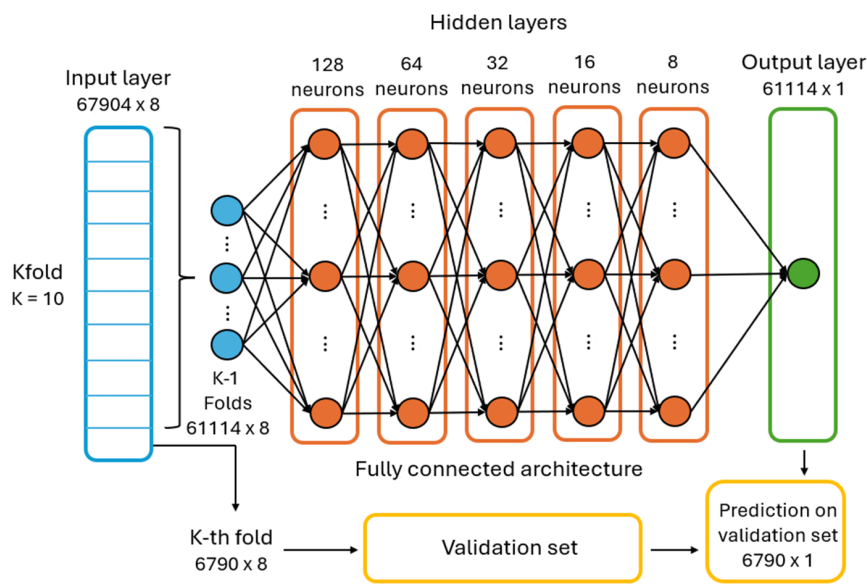

**Fig 2. Architecture of our neural network.** It consists in five hidden layers with progressively decreasing numbers of neurons: 128 64, 32, 16, and 8 neurons in the respective layer. The layers are fully connected. The input layer is divided into ten folds to perform KFold Cross Validation. The blue, orange and green points shown in the architecture represent respectively the input data used for the training procedure, the neurons inside each hidden layer, and the output values, i.e. the predictions.

**Stellar ages: The isochrone method**

**HR diagram.** Individual stellar ages can be predicted by accurately placing the stars in the HR diagram or, equivalently, in the effective temperature - absolute magnitude diagram $(M_G)_0$ (or $M_{G_0}$ to simplify the notation). For these stars, effective temperatures were predicted through our NN model. The subsequent crucial step was to determine accurate values of extinction $A_G$ to derive $M_{G_0}$ using Eq 2. To this aim, we first derived the intrinsic $(G_{BP}-G_{RP})_0$ colors by using our NN effective temperatures and the (inverted) polynomial relation between the Gaia DR3 $(G_{BP}-G_{RP})_0$ colour and the effective temperature, computed as described in Montalto et al. (2021) [64].

From the difference between the observed and intrinsic colors we derived the reddening $E(G_{BP}-G_{RP})$ for all our targets. To convert such reddening into $A_G$, we used the recommended Gaia EDR3 reddening laws, given in the official ESA Gaia documentation, following the method presented in Danielski et al. (2018) [63]. In particular, we constructed a grid by using the equations deriving $k_m = A_m/A_0$ for the bands $m = G_{BP}$ and $G_{RP}$, as a function of the effective temperatures, where $A_0$ is the Gaia reference extinction at 550 nm. Our grid is defined by an array of $A_0$ values ranging from 0.01 to 10.00 with a step size of 0.01 and an array of $T_{eff}$ values ranging from 3 000 to 10 750 K with a step of 250 K. From the combinations of the derived extinctions $A_{BP}$ and $A_{RP}$, we obtained the corresponding values of $E(G_{BP}-G_{RP})$. The extinction values $A_0$ for our targets have been derived by a nearest-neighbour interpolation of the $T_{eff}$ and the $E(G_{BP}-G_{RP})$ values of our targets on this grid. Finally, the extinction $A_G$ has been computed by using the equation $k_m = A_m/A_0$ for the band $m = G$, as a function of the effective temperature.

We derived the $M_{G_0}$ values and constructed the HR diagram using the original $T_{eff}^{GES}$ as well, to compare the results obtained from the NN model predictions with those based on the original spectroscopic effective temperatures.

**Starspots evolutionary models.** Theoretical isochrones used in this work were computed using the Pisa evolutionary code [21,22,82–85]. Among the relevant improvements of these models is the introduction of two additional values of the mixing length parameter $\alpha_{ML}$ equal to 1 and 1.5, to test the effect of the reduced convection efficiency. They are considered alongside the solar-calibrated reference value $\alpha_{ML}$=2. In addition, the models consider the $\beta_{spot}$ parameter, ranging from 0 to 0.6 in step of 0.2, with $\beta_{spot} = 0$ corresponding to the unspotted isochrones, as defined in Franciosini et al. (2022) [22]. Such parameter is the effective spot coverage fraction and take into account both the actual surface spot coverage and the ratio of $T_{spot}/T_{eff}$. The inclusion of star spots in stellar models accounts for their impact on the stellar photosphere. Magnetic fields inhibit convection in these regions, resulting in star spots with temperatures approximately 2 000–3 000 K lower than the effective temperature of the star, depending on the mass. This leads to an effective reduction in the radiating surface area, described by the parameter which quantifies the fraction of the surface covered by spots. To compensate for the blocked energy, models predict an inflation of the stellar radius. The calculation integrates spot coverage into the radiative transfer equations, ensuring self-consistent adjustments of stellar features through iterative feedback until model convergence. This approach provides a physically consistent treatment of star spots in the stellar evolution models.

For the stars in our sample, we considered the models with standard $\alpha_{ML}$=2, since our targets are primarily low mass stars, that are almost adiabatic. In fact, $\alpha_{ML}$ mainly affects the colors of stars with super-adiabatic envelopes. Conversely, we investigated the impact of varying three different values of the spot coverage parameter, $\beta_{spot}$ =0.0, 0.2 and 0.4, as recent studies have shown that this parameter can significantly affect age determinations (e.g. [22,86,87]). In addition, $\beta_{spot} = 0.2$ and 0.4 are consistent with results found from observations of low mass stars (e.g. [22,88]).

Our set of isochrones includes a range of stellar ages spanning from 1 million to 200 million years, except for the set with $\beta_{spot} = 0$ and $ML = 2$, for which we have isochrones down to 100 000 years.

**Stellar isochrone interpolation and starspots model selection.** The ages of the stars are predicted using the isochrone method. Due to the degeneracy introduced by the different $\beta_{spot}$ values, a given star's position on the HR diagram can yield three possible predictions, depending on the chosen $\beta_{spot}$. To resolve this degeneracy, we adopted the following strategy. First, we predicted the three possible ages for each star based on the corresponding $\beta_{spot}$ values.

Next, for each cluster, we computed the median age obtained for each $\beta_{spot}$. Then, we selected the isochrone corresponding to the median age (or to the closest age) and assessed the agreement between this isochrone and the distribution of cluster members in the HR diagram.

To perform the interpolation of each star position in the HR diagram with each set of isochrones, we proceeded in three steps. Firstly, we associated a flag to each isochrone, indicating if the star falls above or under the curve in the HR diagram. To determine the flag, we compared the star's magnitude $M_{G_0}$ with the theoretical magnitude of the isochrones corresponding to the star's temperature $M_{G_{iso}}$. If $M_{G_0} > M_{G_{iso}}$, the star falls under the curve, otherwise it falls above it. In this way, we selected two isochrones, the nearest above and under the star. The next step took into account the computation of the minimum distance between the star and the two selected isochrones. The metric used is the Euclidean distance, since the interpolation is performed in a two-dimensional space and no complex distance metrics are needed. At the end, the star's age is interpolated by a weighted mean, considering the theoretical ages of the two selected isochrones, weighted by the inverse of the minimum distances between the star and the two isochrone curves. This procedure has been applied to each star of the dataset and repeated for each of the three isochrone sets, obtaining three different age predictions for each star.

The subsequent step was to identify the set of isochrones that best represents the position of each stellar cluster on the HR diagram. This required determining the optimal value of the $\beta_{spot}$ parameter that provides the most accurate fit to the observed data. For this purpose, we considered the relationship between $log(T_{eff}^{NN})$ and the magnitude residuals $\Delta M_{G_0}$, computed as the difference between $M_{G_0}$ of cluster members and the magnitudes interpolated from the isochrone corresponding to the cluster's median predicted age. The interpolation uses the star temperature to obtain the expected magnitude based on the isochrone.

A trend in the relationship between $\Delta M_{G_0}$ and $log(T_{eff}^{NN})$ indicates a misalignment between the positions of cluster members and the isochrone, suggesting that the chosen set of isochrones does not accurately represent the cluster. In fact, there is not any physical reason to observe a trend in age as a function of the effective temperature for a stellar cluster.

This analysis was carried out for all three sets of isochrones, each corresponding to a different value of $\beta_{spot}$ and considering both $log(T_{eff}^{GES})$, $log(T_{eff}^{NN})$ and the respective $M_{G_0}$ values. For each set, we calculated three metrics: the slope of the linear regression performed on $\Delta M_{G_0}$ and $log(T_{eff}^{NN})$, which indicates the presence of a possible trend; the absolute sum of the $\Delta M_{G_0}$, providing a measure of overall deviation; the chi-square statistic for $\Delta M_{G_0}$, calculated as the sum of squared residuals divided by the theoretical magnitude value from the isochrone. The best-fitting set of isochrones is identified as the one with the lowest values for these metrics, with the chi-square statistic being the primary criterion.

**Uncertainties on cluster ages.** The procedure used for age prediction is subject to various sources of uncertainty. These include not only stochastic errors of statistical nature but also measurement errors in the distances and in the photometric variables employed to predict the effective temperature. Since the age predictions are derived from the predicted temperatures, any inaccuracies in the photometric measurements propagate through the process, resulting in a spread caused by measurement errors.

The estimation of the error spread was carried out using a Monte Carlo simulation approach. For each star, we randomly generated 100 values for Gaia's G filter magnitude and 100 values for effective temperature, creating 100 fictitious realizations per star. These realizations account for the variability introduced by measurement errors. To generate 100 temperature values for each star, it is essential to define a suitable distribution and specify its parameters. For the temperature, this process is relatively straightforward. Using the bootstrap method previously applied on the Neural Network, we have already obtained an estimate of the standard error for the temperature of each star. Assuming a Gaussian distribution for the simulation procedure, we use the predicted temperature as the mean ($\mu$) and the standard error from the bootstrap as the standard deviation ($\sigma$) of the distribution. With these parameters defined for each star, we can simulate 100 temperature values, ensuring that the variability reflects the uncertainty in the predicted temperatures.

The absolute magnitude $M_G$ in Gaia's G filter is obtained using Eq 2, that depends on the apparent magnitude in the Gaia G filter, the distance $d$ of the star in parsecs, and the extinction value $A_G$. To generate 100 values of $M_G$, we simulated 100 values for each of these three variables, based on their uncertainties. To generate the values for apparent magnitude and distance, we have to specify suitable distributions for these variables. For the apparent magnitude $m_G$, we assume a Gaussian distribution, where the mean $\mu$ is the observed apparent magnitude $m_G$ from Gaia, and the standard error is calculated using the following formula:

$$\sigma_G = \sqrt{\left(\frac{-2.5}{\log(10)} \times \frac{\sigma_{f_G}}{f_G}\right)^2 + (\sigma_{G_0})^2},$$

(3)

where $f_G$ corresponds to the G-band mean flux, $\sigma_{f_G}$ is the associated error and $\sigma_{G_0} = 0.0027553202$ is Gaia EDR3 zero-point uncertainty, given in the official Gaia documentation.

About the distance $d$, a Gaussian distribution is not an appropriate assumption. The distance values adopted in this work [60], provide a lower and upper limits which are not symmetrical. Therefore, we generate distance values using a log-normal distribution, where the expected value is the Gaia photo-geometric distance and the standard deviation is computed as:

$$\sigma_{l_{norm}} = \frac{\log(d_{hi}) - \log(d_{lo})}{2 \times \Phi^{-1}(0.84)},$$

(4)

where $d_{hi}$ and $d_{lo}$ are the upper and lower limits of the distance, respectively, and $\Phi^{-1}(0.84)$ is the inverse cumulative distribution function at the 84th percentile.

For the extinction value $A_G$, we simulated 100 values of temperature of the stars. Then, we calculated the intrinsic color $(G_{BP} - G_{RP})_0$ using the relation adopted in this work [64]. Hence, we calculated the excess $E(G_{BP} - GRP)$ as the difference between the intrinsic color and the observed Gaia DR3 $G_{BP} - G_{RP}$ values. Using the excess $E(G_{BP} - GRP)$, the simulated extinction $A_G$ values were derived using the same procedure for the stars of our dataset and described above.

We generated 100 values of $A_G$ for each star, but it is important to note that the value of $A_G$ is only perturbed if $E(G_{BP} - GRP)$, calculated from the predicted temperature, is greater than or equal to zero. If it is non-physical negative, the extinction value $A_G$ is kept constant, and only the apparent magnitude and distance components are varied to generate the absolute magnitude values. By all these procedure we are able to obtain an estimation of the spread error for each cluster.

For each realization of absolute magnitude and effective temperature, we predicted an age considering the star's position in the HR diagram and using the isochrone interpolation method, applying the appropriate isochrone set for each star, based on the selected $\beta_{spot}$. This process resulted in 100 age values for each star. Using these age fictitious predictions, we computed the standard error associated with the age of each star. To determine the age error spread for the entire cluster, we computed the average of the standard errors across all stars in the cluster. This cluster-level error spread provides a measure of the age dispersion attributable to measurement errors.

## Results

In this section, we present the effective temperature prediction obtained through the Neural Network model trained on the high-quality spectroscopic measurement of the temperature and on the photometric covariates. These predictions are essential for placing stars in the HR diagram and compare their position with state-of-the-art theoretical isochrones to derive accurate stellar ages by isochrone interpolation. To validate our results, we predict stellar ages of selected young clusters for which accurate membership has been derived [53] and compare our results with very recent age determinations of these selected clusters obtained by Jeffries et al. (2023) [18].

**Temperature predictions**

The original dataset of 84 884 stars is divided into training and test set, respectively made up by 67 904 and 16 980 stars, where the test set includes the 20% of the total number of stars of the initial dataset.

In Fig 3, we present the results of the NN model performance on the test set. Temperature predictions obtained through the NN model ($T_{eff}^{NN}$) are compared with the spectroscopic values of the temperature from the GES catalog ($T_{eff}^{GES}$). The figure also shows the residuals of the model ($\Delta T_{eff} = T_{eff}^{GES} - T_{eff}^{NN}$) versus the $T_{eff}^{GES}$ and, on its right, the histogram of the residuals. The colors of the two maps represent the density of the stars, which is computed considering the number of stars that fall in the same area, defined by an hexagon grid, which divides the observed area in hexagon cells and counts the number of points in each cell. For most stars, the model predicts values very close to the reference temperature values of GES. Areas with higher point density are located near the black line, which represents the identity function. In particular, the model performs excellently for stars with low temperatures ($\lesssim 5000$ K). In contrast, for hotter stars, with temperatures above 7000 K, the Neural Network struggles to accurately predict the temperature, tending to obtain lower prediction with respect to the spectroscopic values. To quantify the performance of the model, we computed the mean $\mu$, the standard deviation $\sigma$, the median and the median absolute deviation (MAD) of the residuals $\Delta T_{eff}$. Both the mean and median values indicate a very good accuracy of the predictions, while the standard deviation and the MAD suggest that while the greatest part of predictions are precise, there are some outliers. Therefore, the NN model provides robust temperature predictions for the greatest part of the stars.

The result of the NN approach was compared with benchmark models such as Random Forest (RF) and Support Vector Machines (SVM). The comparison was done considering the MAD computed on the residuals of the three statistical approaches. The NN approach resulted in a MAD value of 160.7, which was lower than both the MAD values of the RF ($MAD_{RF} = 173.67$) and of the SVM ($MAD_{SVM} = 176.15$). This suggests that the NN approach performs better than both RF and SVM on this data.

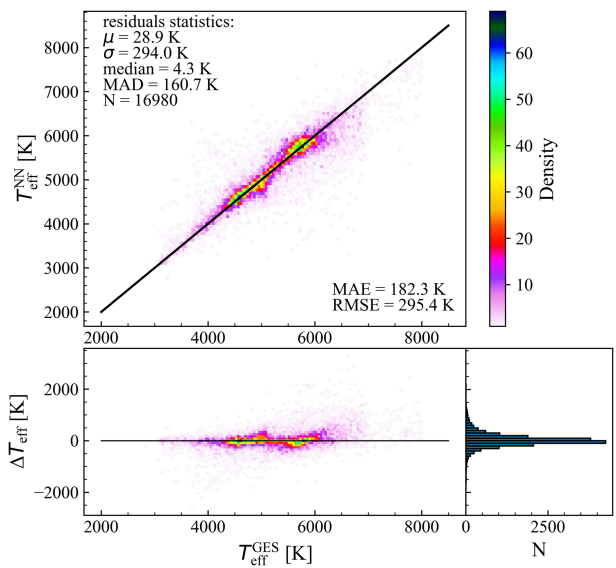

**Fig 3. Effective temperature predictions obtained through NN on the test set of 16 980 stars versus the values of the effective temperature of the GES catalogue.** The black line indicates the identity function (central panel). Residuals of the model ($\Delta T_{eff} = T_{eff}^{GES} - T_{eff}^{NN}$) vs. the $T_{eff}^{GES}$ (bottom panel) and histogram of the residuals (right panel). The color scale is based on the stars' density in the diagram. Residual statistics and prediction errors are reported.

Since the reliability of extinction estimates decreases for stars located beyond 3 kpc, and at the same time our focus is primarily on stars within this distance, we also evaluated the performance of our NN model on the subset of nearby stars with distances <3 kpc, which includes 11 117 objects from the test sample. Focusing on this subsample highlights the improved quality of the predictions for nearby stars. Indeed, the residuals show a mean of 25.2 K, a standard deviation of 283.7 K, a median of 2.2 K, and a MAD of 132.5 K, compared to 28.9 K, 294 K, 4.3 K, and 160.7 K for the full test set. Similarly, both the MAE (164.1 K vs. 182.3 K) and RMSE (284.8 K vs. 295.4 K) are lower for the nearby stars, confirming that the model predictions are more accurate when extinction values are more reliable, even if predictions remain reasonably good even at large distances. In Fig 4, we show the comparison between the NN-predicted temperatures and the spectroscopic GES values for this subsample, together with the residuals and their distribution.

From now on, all the results of the work are based on the prediction obtained through the NN approach.

Once the temperature predictions are obtained, it is necessary to derive the standard error associated with these predictions. To achieve this, we decided to implement a bootstrap procedure based on the trained NN. While the chosen model was carried out using 10-fold cross-validation, for the bootstrap procedure we adopted an early stopping criterion instead of cross-validation in order to reduce computational cost. Importantly, the architecture and hyperparameters of the network were kept fixed. Preliminary tests showed that the results obtained with early stopping were highly consistent with those from cross-validation, thus justifying the use of this more efficient procedure during the bootstrap phase. Specifically, we use the optimised parameters of this NN as the initial parameters for each NN trained during the bootstrap iterations. These parameters are further optimised for each sampled training set, allowing the model to better capture the specific characteristics of each bootstrap sample.

The resulting standard deviations are observed to be lower when the pre-trained neural network is used respect to when it is not employed. At first, this could raise concerns about a potential underestimation of the errors when using the pre-trained model. However, it is important to note that the initial parameters extracted from the pre-trained model are estimated directly from the real data. These parameters represent a starting point that reflects the structure of the dataset, while the standard deviations obtained through bootstrap reflect the variability of the data.

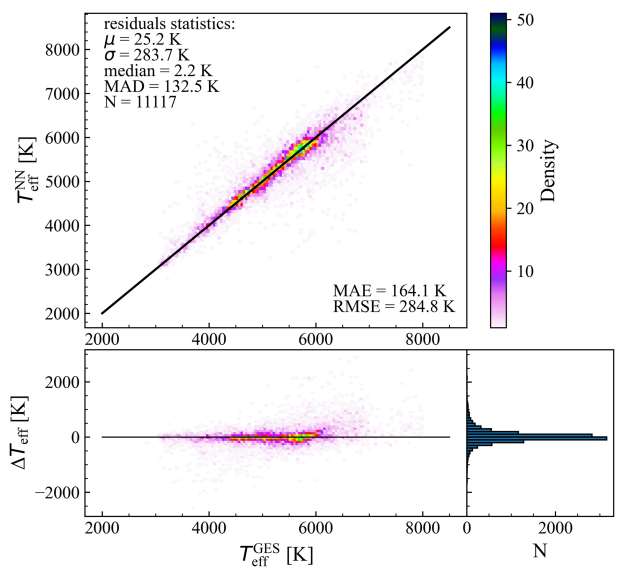

**Fig 4. Effective temperature predictions obtained through NN on the subset of 11 117 stars within 3 kpc, versus the values of the effective temperature from the GES catalogue.** The black line indicates the identity function (central panel). Residuals of the model ($\Delta T_{eff} = T_{eff}^{GES} - T_{eff}^{NN}$) vs. the $T_{eff}^{GES}$ (bottom panel) and histogram of the residuals (right panel). The color scale is based on the stars' density in the diagram. Residual statistics and prediction errors are reported.

Once the temperature predictions are obtained, it is necessary to derive the standard error associated with these predictions. To achieve this, we implemented a bootstrap procedure based on the neural network architecture described above. While the model assessment was carried out using 10-fold cross-validation, for the bootstrap procedure we adopted an early stopping criterion instead of cross-validation in order to reduce computational cost. Importantly, the architecture and hyperparameters of the network (5 layers, 50 epochs, batch size = 8, Adam optimizer, MAE loss) were kept fixed. Preliminary tests showed that the results obtained with early stopping were highly consistent with those from cross-validation, thus justifying the use of this more efficient procedure during the bootstrap phase.

Specifically, the optimised parameters of the trained model were used as the initial parameters for each NN trained during the bootstrap iterations. These parameters were further optimised for each sampled training set, allowing the model to better capture the specific characteristics of each bootstrap sample.

The resulting standard deviations were observed to be lower when the pre-trained neural network was used compared to when it was not employed. At first, this could raise concerns about a potential underestimation of the errors when using the pre-trained model. However, it is important to note that the initial parameters extracted from the pre-trained model are estimated directly from the real data. These parameters represent a starting point that reflects the structure of the dataset, while the standard deviations obtained through bootstrap reflect the variability of the data.

In Fig 5 we show the obtained standard errors as a function of $T_{eff}^{NN}$.

From these results, it is evident that the model performs better for stars with effective temperatures between 4,500 K and 6,500 K, as indicated by the lower values of the standard deviation in this temperature range. This suggests that the NN is more accurate and confident in predicting temperatures for stars within this range, probably because of the higher density of stars and more consistent data available in this region. Higher standard error values are observed for stars at the tails of the temperature distribution, both for cooler stars below 4500 K and hotter stars above 6500 K.

However, the typical uncertainties in the entire range are of the order of 100 K. Overall, the selected NN, with this specific architecture, proves to be a robust statistical method capable of accurately predicting the effective temperature of stars using photometric data extracted from the Gaia and 2MASS surveys. This represents a significant advancement since photometric data can be acquired for stellar samples much larger than spectroscopic ones, and, by applying this approach, we are able to obtain good temperature predictions, which represent one of the most important astrophysical information related to stars.

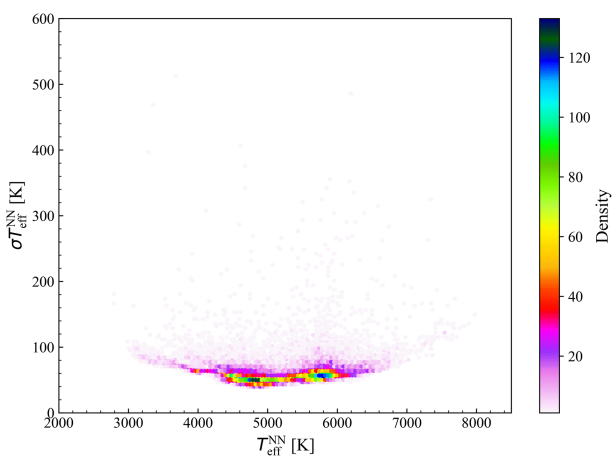

**Fig 5. Comparison of the derived standard error versus the $T_{eff}^{NN}$ of the stars of the test set.** The colors indicate the density map.

**External validation on an independent sample of young stars.** To further assess the robustness and general applicability of our temperature prediction methodology, we performed an external validation on a subset of stars independent from our original study sample. Such external validation is a standard practice in statistical studies, where predictive models are evaluated against representative datasets to ensure that results are not dependent on the initial selection.

For this purpose, we leveraged the astrophysical features from the Gaia DR3 catalog, which represents the most recent, uniform, and comprehensive stellar database currently available in the literature. Although the main Gaia DR3 astrometric and photometric catalog contains more than 1.8 billion stars in total, the effective temperatures, derived from their BP/RP spectra, are available for 471 million sources with G<19, a still very large subset of these stars. Since the focus of our study is on young stars, we extracted the relevant subset by performing a cross-match between the Gaia DR3 table and the Prisinzano et al. 2022 all-sky catalog [28], which contains young stars evenly selected by applying the DBSCAN clustering algorithm to data from the Gaia EDR3 catalog.

From this cross-match, we initially identified 10 857 young stars that are also included in the Gaia DR3 table. For 6 582 of these sources, effective temperatures derived from BP/RP low resolution spectra are available from General Stellar Parameterizer from Photometry (GSP-Phot) Aeneas best library [34]. For these stars, we also considered the extinction in the G band ($A_G$), derived using BP/RP spectra, from GSP-Phot Aeneas best library. In fact, as reported in several studies [34,89], GSP-Phot information, and in particular effective temperatures, are affected by a temperature–extinction degeneracy which causes a simultaneous overestimation of $T_{eff}$ and $A_G$ or, conversely, a simultaneous underestimation of these values. However, this issue mainly arises at higher GSP-Phot extinction values, since for sources with intrinsically low extinction the non-negativity constraint on $A_G$ prevents any significant underestimation [34].

To perform the comparison, we predicted the NN effective temperatures of the matched young stars, by applying the previously chosen NN model. The comparison between the Gaia DR3 effective temperatures ($T_{eff}^{Gaia}$) and the NN predicted temperatures ($T_{eff}^{NN}$) is shown in Fig 6. In order to assess the impact of the Gaia GSP-Phot extinction values, we performed the comparison for the entire matched sample and for a sample restricted to small Gaia GSP-Phot extinction values. In particular, we considered the sample including all 6 582 stars and the one including only stars with GSP-Phot $A_G \leq 2$, that are 5638 in total.

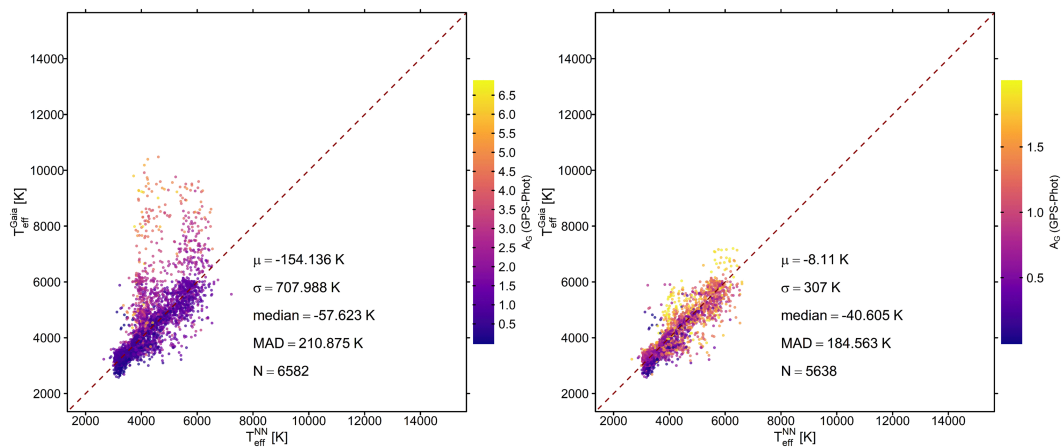

**Fig 6. Comparison of the effective temperature predictions obtained through the NN model and the effective temperature of Gaia on the independent set of young stars.** The dark red line indicates the identity function. The color scale corresponds to the stars' extinction values in the G band ($A_G$). Residual statistics are reported. **Left panel:** all 6 582 stars are included. **Right panel:** only stars with $A_G \leq 2$ are shown, i.e. 5638 objects.

The comparison shows that the NN model generally provides effective temperature predictions consistent with the Gaia DR3 values, with particularly good agreement for stars with low GSP-Phot AG values. Conversely, the largest discrepancies are found for stars with higher GSP-Phot $A_G$ values. For the full sample including stars with high GSP-Phot $A_G$ values, the residuals show a mean difference of $-154$ K, a median difference of $-57$ K, and a median absolute deviation (MAD) of 211 K. Restricting the sample to stars with $A_G \lesssim 2$, the mean difference reduces to 8 K, the median difference to 40 K, and the MAD to 184 K. This external test confirms that the NN model provides reliable temperature predictions for pre-main sequence stars.

### Predicting stellar ages

To test and validate our results, we applied the age interpolation procedure to a list of young clusters ($\lesssim 100-200$ Myr) observed with the Gaia-ESO Survey (GES), for which ages have been derived using model-independent methods. In particular, we considered as reference for the cluster ages, the values obtained in Jeffries et al. (2023) [18], using an empirical model based on the lithium equivalent widths. We selected the first 25 clusters from Table 1 of Jeffries et al. (2023) [18], as their ages fall within our studied age range. We focused only on nearby clusters within 2 kpc from the Sun, since the extinction values can be unreliable beyond this distance. From this list we discarded three clusters: Trumpler 14, whose spectroscopic values are very uncertain due to its large distance (2.35 kpc) [90], NGC 6649 and NGC 6067, for which only a few main-sequence targets have been observed with GES.

Starting from the initial GES dataset of 84 884 stars, which was used to train and test our Neural Network, we considered as members of these 22 clusters those that in Jackson et al. (2022) [53] have cluster membership probability $P_{3D} > 0.9$. Such probability was determined using a maximum likelihood approach based on the 3D kinematics of the targets. From this selection, the sample was reduced to 2 974 cluster members.

Our sample included members that can be equal mass binaries, for which the position on the HR diagram can mimic that of young age PMS stars. In order to exclude them from the analysis, we applied the same methodology illustrated by Valle et al. (2021) [91]. The method is a data-driven rejection technique to select only stars belonging to the single star population from the absolute magnitude-temperature diagram. We used this approach for Blanco1, NGC 2516, NGC 6709, 25 Ori, Gamma Velorum, IC 2602, NGC 2232, NGC 2547 and rho Ophiuchi. Through this selection, we identified 172 binary stars belonging to these clusters, which were excluded from the analysis, reducing our dataset to 2 802 entries.

Additionally, we excluded stars with temperatures above 7 000 K, as, based on the results from the Neural Network, the model struggles to accurately predict the temperature of such stars. Moreover, these hotter stars in the GES catalog are associated with high measurement errors in the effective temperatures.

Our final sample includes 2 679 stars, belonging to 22 different clusters.

**Starspots model selection.** For each cluster of our sample, we derived three sets of individual stellar ages using the isochrone method and the PISA solar metallicity isochrones for the three $\beta_{spot}$ values 0.0, 0.2 and 0.4. Then, we performed the $\beta_{spot}$ selection procedure, described in *Data and Methods*, to find the best isochrone set for each cluster. As an example, in Fig 7, we present the results of this procedure for the cluster 25 Ori. The plots clearly indicate that, for this cluster, the best $\beta_{spot}$ value is 0.4, as the magnitude residuals $\Delta M_{G_0}$ do not exhibit any discernible trend, with the slope being the closest to zero among the three $\beta_{spot}$ values. Furthermore, the chi-square statistic reaches its minimum value for $\beta_{spot} = 0.4$, confirming that this is the best-fitting isochrone set for the cluster, as the stars in the HR diagram align more closely.

In Table 1, we present the selected isochrone set for each cluster, obtained using the spectroscopic and the predicted effective temperatures by the NN approach, that we indicate with $\beta_{spot}^{spec}$ and $\beta_{spot}^{NN}$, respectively.

For 18 of the 22 clusters we derive the same value, (i.e. $\beta_{spot}^{spec} = \beta_{spot}^{NN}$), confirming a very good agreement between spectroscopic and predicted effective temperatures, reproducing very similar HR diagrams. A different $\beta_{spot}$ value is found for the very young cluster NGC 6530, for which our metrics gives very similar values by adopting $\beta_{spot} = 0.2$ and 0.4.

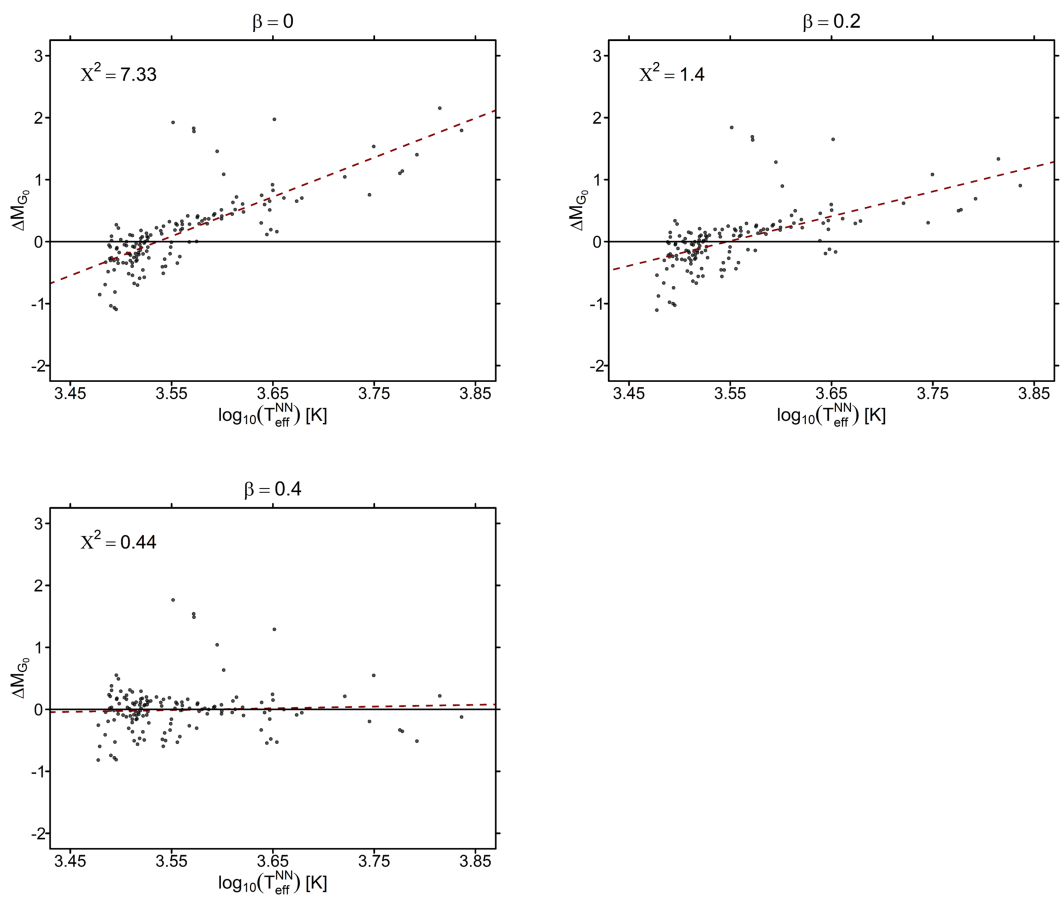

**Fig 7. Comparison of the residuals of the magnitude G versus the effective temperature for the three isochrone sets with for the 25 Ori cluster.** The black line indicates the zero value, the dashed line indicates the linear trend of the relationship. In the top left of each graph, the corresponding chi-square statistic is reported.

This is due to the absence of a well-defined sequence and the significant luminosity spread in the HR diagram of this cluster, which prevents us from definitively constraining the $\beta_{spot}$ value. We retrieved different $\beta_{spot}$ values also for the older clusters NGC 2451a and NGC 2451b. In this case, the discrepancy may be attributed to a poor fit caused by the limited number of cluster members. Finally, in the case of NGC 2516 the median cluster age obtained for $\beta_{spot}^{NN} = 0.4$ is likely affected by the upper limit (200 Myr) of our isochrone set. A significantly better fit is obtained for NGC 2516 by adopting $\beta_{spot} = 0.2$, using both spectroscopic and NN model effective temperatures.

Our results clearly show that, for all the star clusters included in the analysis, a $\beta_{spot}$ value greater than zero is required to accurately describe the predicted positions of the stars on the HR diagram. In particular, for the youngest clusters, the most suitable set of isochrones is that with $\beta_{spot} = 0.4$, while for the oldest one we retrieve $\beta_{spot} = 0.2$ as best value, with the exception of the cluster NGC 6709. We note, however, that for this latter cluster, the fit is quite poor since it is significantly limited by the low statistics (only 39 members) and the small dynamical range of the main sequence covered by GES observations ($M_{G_0} \lesssim 7$).

The age trend of $\beta_{spot}$ that we observe with our procedure is consistent with theoretical predictions, as younger and more magnetically active clusters are expected to exhibit higher spot coverage. We note that, compared to $\beta_{spot}^{NN}$, such trend is more stable and well defined for $\beta_{spot}^{spec}$, as spectroscopic effective temperatures are subject to lower

**Table 1. Results of the cluster age analysis.** Column 3 lists the number of members of each cluster, while column 4 provides the best-fit $\beta$ values derived in this work based on spectroscopic effective temperatures. Columns 5 and 6 present the mean age and its standard deviation, both expressed on a logarithmic scale, while column 7 reports the contribution of age spread due to the combination of the statistical measurement errors, obtained from Monte Carlo simulations, and the systematic error 0.1 due to unresolved binaries. Columns 8 to 10 report the ages corresponding to the 50th (median), 16th and 84th percentiles, respectively. Column 1 provides the short labels (A, B, C, ...) used in the figures.

| Label | Cluster | N | $\beta$ | Mean Log Age [yr] | $\sigma_t$ [dex] | $\sigma_e$ [dex] | Median Log Age [yr] | Log Q16 [yr] | Log Q84 [yr] |
|---|---|---|---|---|---|---|---|---|---|
| A | NGC 6530 | 325 | 0.4 | $6.55 \pm 0.02$ | 0.41 | 0.24 | 6.51 | 6.09 | 6.96 |
| B | Chamaeleon I | 76 | 0.4 | $6.80 \pm 0.04$ | 0.36 | 0.27 | 6.78 | 6.42 | 7.13 |
| C | Rho Ophiuchus | 36 | 0.4 | $7.02 \pm 0.06$ | 0.37 | 0.48 | 6.96 | 6.75 | 7.24 |
| D | NGC 2264 | 471 | 0.4 | $6.93 \pm 0.02$ | 0.43 | 0.35 | 6.98 | 6.51 | 7.26 |
| E | NGC 2244 | 112 | 0.4 | $6.66 \pm 0.04$ | 0.37 | 0.22 | 6.66 | 6.32 | 6.95 |
| F | Lambda Ori | 179 | 0.4 | $6.98 \pm 0.03$ | 0.38 | 0.32 | 7.01 | 6.61 | 7.28 |
| G | Lambda Ori B35 | 43 | 0.4 | $6.77 \pm 0.06$ | 0.36 | 0.31 | 6.82 | 6.36 | 7.15 |
| H | 25 Ori | 145 | 0.4 | $7.32 \pm 0.02$ | 0.27 | 0.27 | 7.32 | 7.09 | 7.46 |
| I | ASCC 50 | 175 | 0.4 | $6.76 \pm 0.03$ | 0.37 | 0.30 | 6.81 | 6.35 | 7.09 |
| J | Collinder 197 | 98 | 0.4 | $6.69 \pm 0.04$ | 0.35 | 0.26 | 6.74 | 6.31 | 7.06 |
| K | Gamma Velorum | 162 | 0.4 | $7.30 \pm 0.01$ | 0.17 | 0.30 | 7.31 | 7.20 | 7.41 |
| L | IC 4665 | 31 | 0.4 | $7.82 \pm 0.06$ | 0.35 | 0.29 | 7.80 | 7.46 | 8.30 |
| M | NGC 2232 | 44 | 0.4 | $7.74 \pm 0.03$ | 0.22 | 0.27 | 7.69 | 7.61 | 7.88 |
| N | NGC 2547 | 107 | 0.4 | $7.58 \pm 0.02$ | 0.25 | 0.36 | 7.56 | 7.42 | 7.65 |
| O | IC 2602 | 48 | 0.2 | $7.48 \pm 0.04$ | 0.25 | 0.34 | 7.56 | 7.19 | 7.65 |
| P | NGC 2451b | 56 | 0.2 | $7.54 \pm 0.03$ | 0.24 | 0.34 | 7.54 | 7.40 | 7.62 |
| Q | IC 2391 | 31 | 0.2 | $7.54 \pm 0.04$ | 0.22 | 0.40 | 7.57 | 7.28 | 7.76 |
| R | NGC 2451a | 38 | 0.2 | $7.62 \pm 0.05$ | 0.29 | 0.26 | 7.64 | 7.32 | 7.75 |
| S | NGC 6405 | 49 | 0.2 | $7.57 \pm 0.04$ | 0.28 | 0.23 | 7.64 | 7.33 | 7.78 |
| T | NGC 2516 | 332 | 0.2 | $7.93 \pm 0.02$ | 0.28 | 0.33 | 7.96 | 7.61 | 8.30 |
| U | Blanco 1 | 83 | 0.2 | $8.08 \pm 0.02$ | 0.21 | 0.32 | 8.07 | 7.92 | 8.30 |
| V | NGC 6709 | 39 | 0.2 | $7.91 \pm 0.08$ | 0.48 | 0.35 | 7.97 | 7.55 | 8.30 |

https://doi.org/10.1371/journal.pone.0336592.t001

uncertainties than those derived from the NN model. For this reason, from here on we predict the cluster ages assuming the $\beta_{spot}^{spec}$ values, based on the spectroscopic effective temperatures, and given in Table 1.

**Cluster ages.** Fig 8 shows the HR diagram of the 22 clusters, obtained using the predicted effective temperatures by the NN and the comparison with the three set of isochrones, corresponding to each median cluster age. The median age isochrone obtained with the best $\beta_{spot}$ is highlighted.

Fig 9 shows the box plots of the predicted age distributions of the 22 young stellar clusters, derived using the most suitable starspots evolutionary models. Table 1 shows the different statistical measures of the cluster age distributions along with the total stellar age spread, $\sigma_t$, computed as the standard deviation of the age distributions. Additionally, it includes $\sigma_e$, which represents the contribution of age spread due to the combination of the statistical measurement errors, obtained from Monte Carlo simulations and the systematic error 0.1 due to unresolved binaries. We found that, for all clusters, with the exception of Rho Ophiuchi and IC 2391 which have a number of members <50, the uncertainties $\sigma_e$ are within $\pm 0.35$ dex due to observational errors.

To assess the effectiveness of our NN model in predicting temperatures and consequently ages, compared to spectroscopic temperatures, we performed the comparison between the mean cluster ages, predicted using the NN effective temperatures, and those predicted using the original spectroscopic GES effective temperatures, as shown in Fig 10. The error bars represent the error on the mean ages, computed as $\sigma_t / \sqrt{(N)}$, where N is the number of stars used to compute the mean, for the two sets of effective temperatures. This comparison demonstrates a very good agreement, with a mean difference in ages of -0.005 dex and a standard deviation of 0.067 dex. These results indicate that adopting our NN effective temperatures, we achieve the same level of accuracy and precision as using the original spectroscopic effective temperatures.

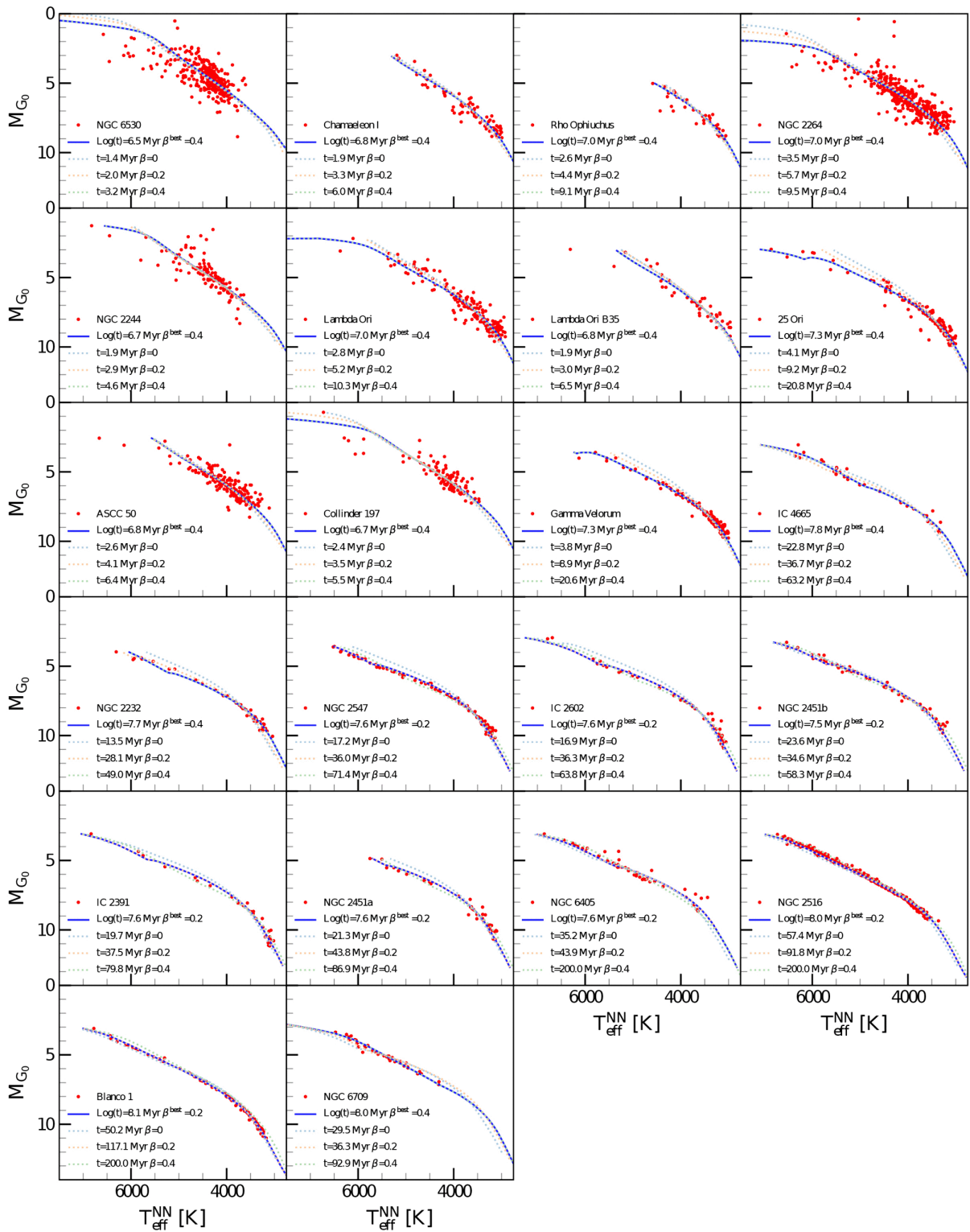

**Fig 8. HR diagram for the 22 selected clusters.** Red symbols represent the selected cluster members [53], with candidate binaries removed. The blue solid line corresponds to the solar metallicity PISA isochrone for the best $\beta_{spot}$ value (see text) at the median cluster age reported in the legend. The dotted lines show isochrones at the median cluster age computed using stellar ages derived from the isochrone sets with the three possible $\beta_{spot}$ values.

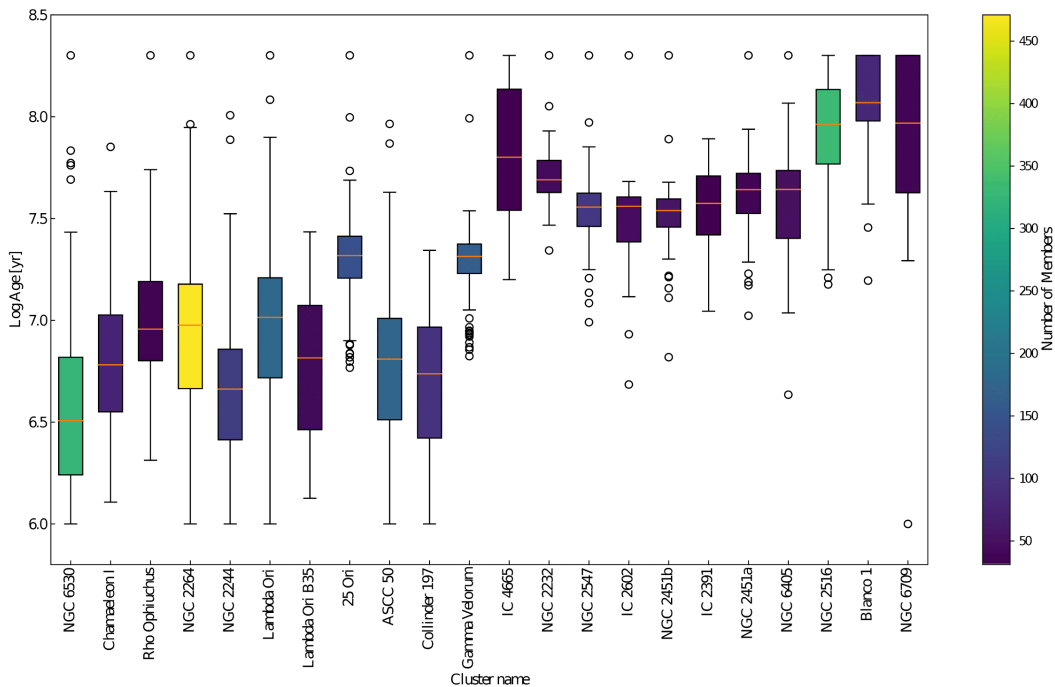

**Fig 9. Box plots of the predicted age distributions of the 22 young stellar clusters.** The x-axis lists the clusters, and the y-axis shows the corresponding age distributions. The boxes indicate the interquartile range (IQR) between the first and third quartiles (Q1 and Q3), while the line inside the box represents the median value. The whiskers extend to the minimum and maximum values within 1.5 times the IQR, and the circles represent outliers, values that lie outside the range of the whiskers. The colors of the box plots represent the number of members in each cluster.

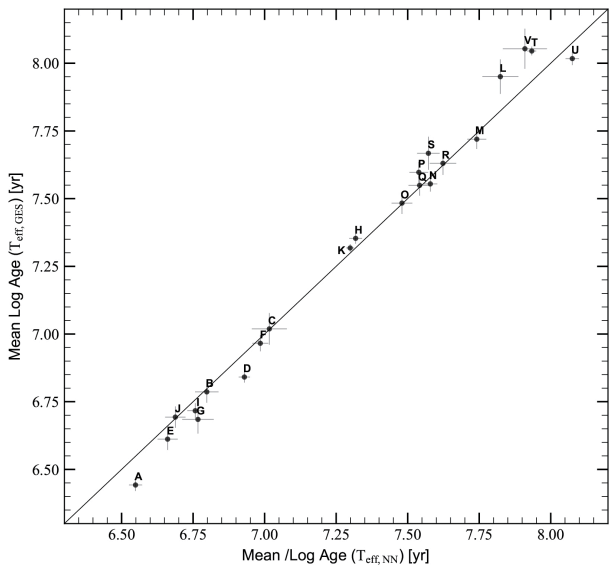

**Fig 10. Mean cluster ages derived from GES spectroscopic effective temperatures versus those obtained from NN effective temperatures.** Error bars are the standard deviations of the distributions.

Finally, as external check, we compared our results with the values used in Jeffries et al. (2023) [18] as training ages and those obtained from the best-fit lithium ages. The comparisons, shown in Figs 11 and 12, highlights generally good agreement. With respect to the training ages, we observe a mean difference of 0.014 dex, and a standard deviation of 0.248 dex, with a t-statistic of 0.258 and a p-value of 0.799. With respect to the best-fit lithium ages, the mean difference

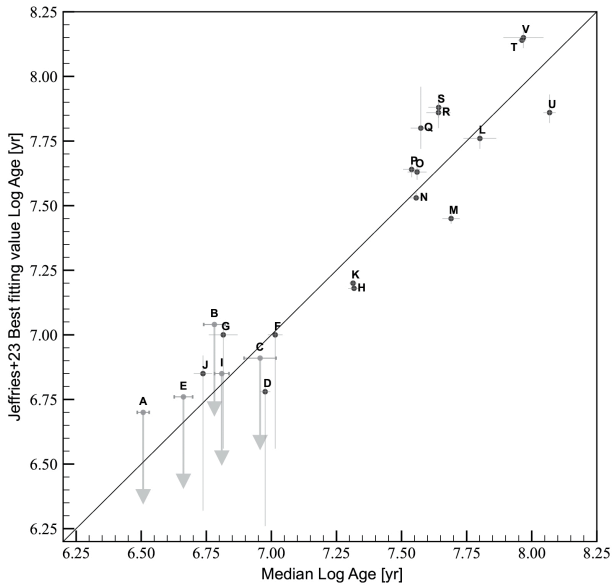

**Fig 11. Comparison of training ages given in [18] with cluster ages predicted in this work.** Error bars are the standard deviations of the stellar age distributions. Gray arrows show ages of clusters where only a 95 per cent upper limit has been determined in Jeffries et al. [18]

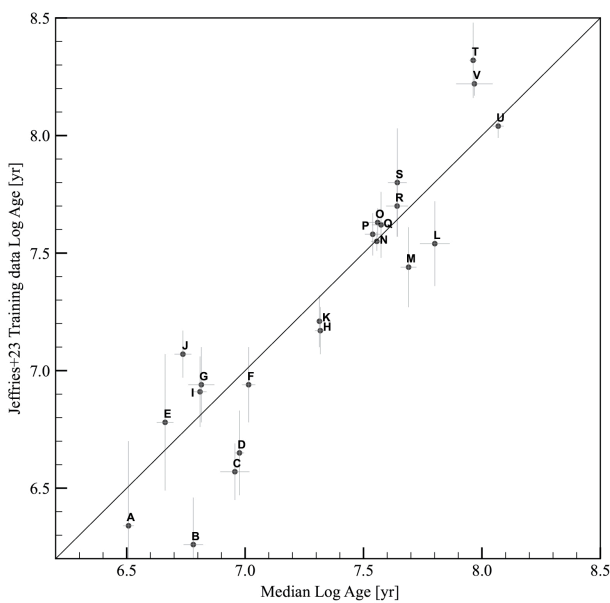

**Fig 12. Comparison of best-fitting lithium ages given in [18] with cluster ages predicted in this work.** Error bars are the standard deviations of the stellar age distributions. Gray arrows show ages of clusters where only a 95 per cent upper limit has been determined in Jeffries et al. [18]

is –0.053 dex and the standard deviation is 0.183 dex, with a t-statistic of –1.160 and a p-value of 0.263, suggesting that the differences between our predicted ages and those from the literature are statistically negligible. The most discrepant clusters are the youngest ones, namely NGC 6530, Rho-Ophiuchi, NGC 2264, and Cha I, for which our predicted ages using isochrones with starspots are higher than the typical ages reported in the literature for these clusters. However, the agreement improves when considering the best-fit lithium ages, although Jeffries et al. determined only a 95 percent upper limit for the ages of these young clusters. Discrepancies are observed for the clusters NGC 2516 and NGC 6709, for which our results are limited by the small coverage of the cluster sequence of the members, particularly in the high-mass range, which is crucial for constraining the ages of clusters older than 100 Myr.

## Discussion

The Neural Network-based machine learning algorithm developed in this work, trained on spectroscopic effective temperature measurements, enables accurate predictions of effective temperatures using only photometric and astrometric data. This method provides effective temperatures with an accuracy comparable to that of spectroscopic data, particularly for stars within 3 kpc, where extinction uncertainties are lower.

The most important advantage of this approach is its ability to construct HR diagrams for young stellar clusters (age < 200 Myr) using extensive photometric datasets, even in cases where obtaining spectroscopic measurements is challenging or impossible. Accurate HR diagrams are essential for reliably determining individual stellar masses and ages using the evolutionary models.

The adoption of state-of-the-art isochrones that include starspots effects, through the parameter $\beta_{spot}$, which takes into account the fraction of stellar surfaces covered by spots and their temperatures, represents a further significant step forward in deriving reliable stellar ages. The predicted cluster ages show good agreement with those reported very recently in the literature [18] and determined with a different method, underscoring the reliability of our method. Nonetheless, discrepancies are observed for some of the youngest clusters, such as NGC 6530, NGC 2264, Chameleon I and Rho-Ophiuchi. For these clusters, our starspot-based isochrones predict ages up to a factor of three older than the ages derived in the literature(e.g. [23,92,93]). Similarly, the ages derived using starspot-based isochrones are up to a factor of three older than those obtained adopting nonspotted models, as shown in Fig 8. This result may seem unexpected, given that Class 0/I young stellar objects (protostars) have been identified in these clusters. However, we emphasize that the ages reported in this work refer specifically to optically visible PMS stars (Class II and III) observed with Gaia DR3. Therefore, we cannot rule out the possibility that these clusters may also contain stars younger than the mean cluster age we estimated.

We note that in the case of the oldest clusters in our sample, namely NGC 2516 and NGC 6709, the limited extent of the stellar sequence toward higher masses reduces the accuracy of the ages derived with our method. This is because, at ages older than 100 Myr, isochrones become tightly packed, making age determination more challenging. Although these findings highlight the importance of a complete cluster sequence for accurately constraining the ages of clusters older than 100 Myr, we emphasise that the primary aim of this work is to derive accurate ages for clusters younger than 50–100 Myr.

Our findings on the youngest clusters are fully consistent with conclusions found by Somers et al. (2020) [94]. They also found that models incorporating starspots predict older ages compared to standard, spotless models. This is because starspots modify the stellar structure by reducing the effective temperature and increasing the radius. As a result, standard isochrones underestimate the ages of spotted stars, leading to systematic age discrepancies when comparing models with and without starspots.

Although our sample does not include high mass stars ($> 2\,M_\odot$), the adoption of these models allows us to figure out the long-standing issue in the literature of the mass-dependent spread in the inferred ages of PMS stars, where low-mass stars, particularly M stars, appear systematically younger than higher-mass F-type stars within the same

cluster—sometimes by up to a factor of two [95,96]. Similar conclusions have been drawn by Somers et al.(2020) [94], who identified a specific $f_{spot}$ value in Upper Sco at which stars of different masses (above and below $0.9\,M_\odot$) converge to the same average age. This critical $f_{spot}$ value, approximately 0.5, is considered the value at which the mass-dependent age discrepancy is minimized.

This assumption conceptually aligns with our method for determining $\beta_{spot}$ for which we assume that all stars corresponding to a given isochrone—compatible with the cluster's median age computed for three possible $\beta_{spot}$ — do not exhibit a systematic trend in magnitude as a function of effective temperatures. While this assumption requires further studies to test the dependence of magnetic activity on stellar mass, the resulting cluster ages obtained under this framework are consistent with other model-independent methods. For example, it has been found that cluster ages derived from lithium depletion patterns, are at least a factor of two older than previously predicted [18,22,97]. Moreover, the ages of high-mass stars, which have radiative envelopes, are expected to be not affected by spotted models [98], whereas low-mass stars are significantly impacted, as shown in this work and in other studies adopting spotted models [94,98]. This suggests that our assumption is valid on average, as our sample is dominated by low-mass stars, where the effects of starspots are most significant.

The adoption of isochrones with $\beta_{spot} > 0$ across our cluster sample indicates that evolutionary models incorporating starspots provide a more accurate description of stellar properties. The observed trend of $\beta_{spot}$ with age — $\beta_{spot} = 0.4$ being most suitable for clusters younger than 10–30 Myr, and $\beta_{spot} = 0.2$ being more appropriate for older clusters — suggests that the influence of starspots diminishes as stars evolve. This trend confirms previous findings about the early stages of stellar evolution, highlighting the role of magnetic activity in shaping the observable properties of young stars, as very recently found by Pérez Paolino et al. (2025) [99].

Previous findings shows that using the HR diagram as a tool for inferring individual stellar ages involves dealing with the non-trivial challenge of accurately interpreting stellar observables within the framework of stellar structure and atmosphere theory, particularly for low-mass young stars characterised by highly complex physics. This means testing evolutionary models of very young stars. But, deriving accurate individual stellar ages with the HR diagram method has an additional valuable outcome which is gaining insights into the processes by which stars form within stellar clusters, using the age spreads in SFRs as a direct observational tool to constrain the debated star formation timescale.

For several decades, numerous SFRs have been observed to exhibit an apparent spread in luminosity. This spread has been interpreted by various authors as evidence of age spread (e.g. [9,23,98]), while others have attributed it to uncertainties in the observational and astrophysical features used to derive stellar ages(e.g. [95]). The sample of SFRs and old open clusters analysed in this work, for which stellar ages have been consistently derived using homogeneous photometric and spectroscopic data alongside the same methods and theoretical models, provides an ideal dataset to address this question.

In agreement with what is typically observed in the literature, the HR diagrams in our sample also exhibit a luminosity spread that decreases significantly after the first 10–20 Myr (see Fig 8). To investigate the nature of the luminosity spread observed in the clusters younger than 10-20 Myr, we compared the predicted total observed spread, $\sigma_t$, computed as the standard deviation of stellar ages, with that expected from observational uncertainties, $\sigma_e$. The latter was obtained by quadratically combining the statistical errors derived from the mean of the MC simulated stellar ages, with a systematic error of 0.1 dex, accounting for the presence of unresolved binaries [23]. Then, we defined the intrinsic age spread as

$$\sigma_i = \sqrt{\sigma_t^2 - \sigma_e^2}, \tag{5}$$

and the associated uncertainty as

$$\delta\sigma_i = \sqrt{(\delta\sigma_t)^2 + (\delta\sigma_e)^2}, \tag{6}$$

where

$$\delta\sigma_t = \frac{\sigma_t}{\sqrt{2N}}, \tag{7}$$

and $\delta\sigma_e$ is the standard deviation of the uncertainties associated with the simulated stellar ages.

The results for the clusters for which $\sigma_t > \sigma_e$ are given in Table 2 and shown in Fig 13. For these clusters, $\sigma_i$ exceeds the uncertainties by at least $1\,\sigma$, with more significant results observed for the clusters NGC 6530, NGC 2244, Chameleon I, and Collinder 197, where $\sigma_i$ is larger by more than $2\,\sigma$. Then, for these clusters, $\sigma_i$ can be attributed to a real and intrinsic age spread, confirming the hypothesis of a slow SF process, during which star formation consists of subsequent events. These results would support the hypothesis of SF occurring in multiple events over time and it is in agreement with previous findings in the literature(e.g. [23,98,100,101]) which provide evidence of more than one epoch of SF.

## Conclusions

In this study, we applied and validated a Neural Network model to accurately predict the effective temperatures of pre-main sequence (PMS) stars using photometric data from the Gaia DR3 and 2MASS catalogs. This approach enables the prediction of accurate stellar characteristics for a much larger population of stars than those accessible through spectroscopic observations alone. The application of the Neural Network model, trained on high-quality spectroscopic temperature measurements extracted from the GES catalog, proved to be highly effective in capturing the complex relationships between photometric variables and spectroscopic temperatures, achieving robust and precise predictions. Through validation procedures and comparison with benchmark methods such as Random Forest and Support Vector Machines, the Neural Network consistently confirmed to be a valid approach for this purpose, outperforming the predictive performance of the other models. The integration of bootstrap techniques further allowed the derivation of prediction uncertainties, ensuring a robust quantification of errors, with particularly reliable results for stars within 3 kpc, which are also the main focus of this work. Indeed, our interest is primarily directed toward nearby PMS stars, and we therefore concentrated on achieving optimal predictions for this population.

The predicted temperatures were successfully employed to place stars on the Hertzsprung-Russell diagram, leading to accurate age predictions for selected young stellar clusters using the isochrone interpolation method. The results show excellent agreement with ages derived for the same clusters by considering the spectroscopic values of the temperature, as well as with recent literature age values based on model-independent techniques like lithium equivalent widths. These results confirm the goodness of the predictions obtained by the proposed Neural Network, pointing out how, basing exclusively on photometric information, we are able to achieve excellent results in predicting the temperatures of these young stars, particularly for temperatures below 7 000 K, which corresponds to the range where low-mass young stars, the main focus of this work, are typically found. Furthermore, regarding age prediction, the inclusion of starspot evolutionary models helped to obtain more consistent results, demonstrating that for all the stars clusters considered in the analysis, a $\beta_{spot}$ value greater than zero is necessary to accurately represent the stars' positions on the HR diagram, indicating that evolutionary models incorporating starspots provide a more accurate description of stellar properties. Specifically, the isochrones with $\beta_{spot} = 0.4$ were found to be the most appropriate for the youngest clusters, i.e. clusters younger than 10-30 Myr, while for the oldest ones, i.e. cluster with predicted age between 50 and 100 Myr, $\beta_{spot} = 0.2$ emerged as the optimal choice. The observed age-dependent trend in $\beta_{spot}$ aligns with theoretical expectations, as younger and more magnetically active clusters are predicted to exhibit higher levels of spot coverage.

Table 2. **Clusters with evidence of age spread.** Columns 2 to 10 give the logarithmic mean age, the total age spread ($\sigma_t$), the error associated on the total age spread ($\delta\sigma_t$), the spread due to observational errors ($\sigma_e$), the uncertainty on the spread due to observation errors ($\delta\sigma_e$), the intrinsic age spread ($\sigma_i$) and its corresponding error ($\delta\sigma_i$). Dispersions are given in dex.

| Label | Cluster | Mean Log Age [yr] | $\sigma_t$ | $\delta\sigma_t$ | $\sigma_e$ | $\delta\sigma_e$ | $\sigma_i$ | $\delta\sigma_i$ | $\delta\sigma_i$ |
|-------|---------|-------------------|-----------|------------------|-----------|------------------|-----------|------------------|------------------|
| A | NGC 6530 | $6.55 \pm 0.02$ | 0.41 | 0.02 | 0.24 | 0.11 | 0.33 | 0.11 | 0.11 |
| B | Chamaeleon I | $6.80 \pm 0.04$ | 0.36 | 0.03 | 0.27 | 0.09 | 0.23 | 0.10 | 0.10 |
| D | NGC 2264 | $6.93 \pm 0.02$ | 0.43 | 0.01 | 0.35 | 0.15 | 0.25 | 0.15 | 0.15 |
| E | NGC 2244 | $6.66 \pm 0.04$ | 0.37 | 0.02 | 0.22 | 0.09 | 0.30 | 0.10 | 0.10 |
| F | Lambda Ori | $6.98 \pm 0.03$ | 0.38 | 0.02 | 0.32 | 0.11 | 0.20 | 0.12 | 0.12 |
| G | Lambda Ori B35 | $6.77 \pm 0.06$ | 0.36 | 0.04 | 0.31 | 0.11 | 0.19 | 0.12 | 0.12 |
| I | ASCC 50 | $6.76 \pm 0.03$ | 0.37 | 0.02 | 0.30 | 0.11 | 0.21 | 0.11 | 0.11 |
| J | Collinder 197 | $6.69 \pm 0.04$ | 0.35 | 0.03 | 0.26 | 0.09 | 0.24 | 0.09 | 0.09 |

Finally, this study investigated the intrinsic age spread within young stellar clusters, revealing important insights into their formation and evolution. By accounting for both observational errors and systematic uncertainties, we estimate the intrinsic spread of stellar ages for each cluster as the difference between the total age spread and the measurement age spread of them. The results show that, while measurement and observational errors contribute significantly to the total age spread, an intrinsic age spread is still evident for some clusters, especially for the youngest. This result can be an additional evidence to find an answer to the discussion in astrophysical research concerning the duration of the formation process of star clusters, supporting the theoretical astrophysical models which suggest multiple formation events during the star formation process.

The methodology proposed in this work for estimating effective temperatures and fundamental stellar properties, such as ages and masses, is expected to be valuable in the near future with the arrival of large spectroscopic surveys like 4MOST [102] and WEAVE [20]. As these surveys will analyse tens of millions of spectra and provide more accurate spectroscopic temperatures, we will use this data to improve our method by expanding the training sample, thus enabling more reliable temperature predictions also for stars located beyond 3 kpc, where current extinction-related values are less reliable.

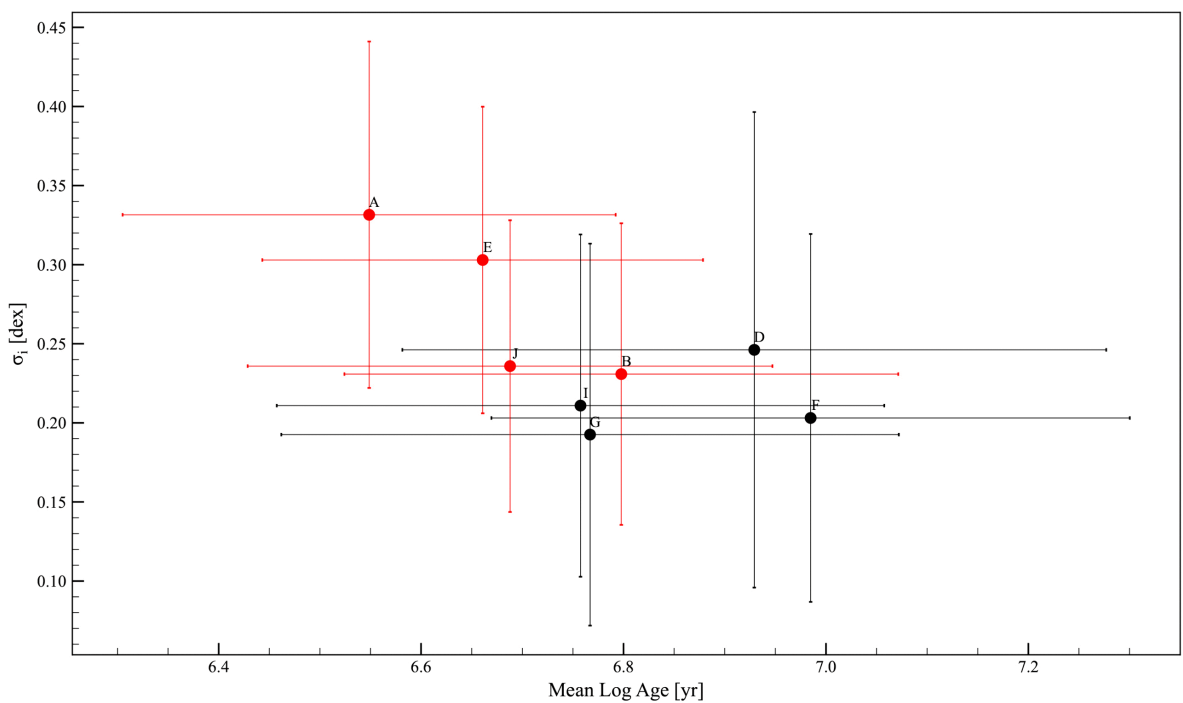

**Fig 13. Intrinsic age spread, computed as described in the text, for the youngest cluster of our sample for which such spread is significantly larger than the expected errors within 1 (black symbols) or 2 $\sigma$ (red symbols).**

Applying this methodology to large samples of stars with only photometric data available will enable us to address key astrophysical questions, such as the Star Formation History and the Initial Mass Function of complex star-forming regions on a Galactic scale.

## Acknowledgments

The research work of Nicoletta D'Angelo and Giada Adelfio was supported by the Targeted Research Funds 2025 (FFR 2025) of the University of Palermo (Italy) and the European Union - NextGenerationEU, Mission 4, Component 2, in the framework of the GRINS -Growing Resilient, INclusive and Sustainable project (GRINS PE00000018 – CUP C93C22005270001).

MUR- PRIN 2022: Spatio-temporal Functional Marked Point Processes for probabilistic forecasting of earthquakes 2022BN7CJP. P. I. Giada Adelfio. CUP B53C24006340006; National Recovery and Resilience Plan (NRRP), Mission 4 Component 2 Investment 1.4 - Call for tender No. 3138 of 16 December 2021, rectified by Decree n.3175 of 18 December 2021 of Italian Ministry of University and Research funded by the European Union – NextGenerationEU, Award Number: Project code $CN$–00000033, Concession Decree No. 1034 of 17 June 2022 adopted by the Italian Ministry of University and Research, CUP UNIPA B73C22000790001, Project title "National Biodiversity Future Center - NBFC".

We thank Marco Montalto for providing the updated version of the polynomial relation between the colours $(G_{BP} - G_{RP})_0$ and the effective temperatures, based on *Gaia* DR3 data. We also thank Emanuele Tognelli for providing state-of-the-art starspot-dependent models and for the insightful discussions on the physical processes integrated into them.

## Author contributions

**Conceptualization:** Loredana Prisinzano.

**Data curation:** Marco Tarantino.

**Formal analysis:** Marco Tarantino.

**Funding acquisition:** Giada Adelfio.

**Investigation:** Loredana Prisinzano.

**Methodology:** Marco Tarantino, Loredana Prisinzano.

**Software:** Marco Tarantino.

**Supervision:** Loredana Prisinzano, Giada Adelfio.

**Validation:** Marco Tarantino, Francesco Damiani.

**Visualization:** Marco Tarantino, Loredana Prisinzano.

**Writing – original draft:** Marco Tarantino, Loredana Prisinzano.

**Writing – review & editing:** Marco Tarantino, Loredana Prisinzano, Nicoletta D'Angelo, Francesco Damiani, Giada Adelfio.